\documentclass{amsart}
\newtheorem{theorem}{Theorem}
\begin{document}
\sloppypar
\author{Ilja Schmelzer}
\title{A doubler-free lattice theory for QCD based on geometric fermions}

\begin{abstract}

We present doubler-free gauge-invariant lattice vector gauge action for some real representations of Wilson gauge fields on an octet of fermions.  It is based on a geometric representation of the Dirac equation as an evolution equation on the three-dimensional exterior bundle $\Lambda(\mathbb{R}^3)$ for a single bispinor and of the bundle $(\Lambda\otimes\Lambda)(\mathbb{R}^3)$ for an octet.  We find doubler-free lattice Dirac operators for above bundles.  A gauge-invariant connection with Wilson lattice gauge fields is possible for some real representations of the gauge group.  The QCD action of $SU(3)$ is of this type.

Application in lattice QCD seems useful: We don't have to waste time and memory for doublers as well as for correction terms to suppress them.

\end{abstract}

\maketitle

\section{Introduction}

One of the main problems of lattice gauge theory is the fermion doubling problem.  If we use a naive central differences discretization of the Dirac equation, we observe a doubling effect: The large distance limit of the resulting lattice equation does not describe the original single Dirac particle, but, instead, sixteen Dirac particles --- a factor two doubling in each spacetime dimension.  Such a discretization is inappropriate for quantum computations where the number of degrees of freedom is important.

A partial solution is given by a geometric variant of the Dirac equation --- the Dirac-K\"ahler equation \cite{Kaehler}.  This equation is defined on the complexified four-dimensional exterior bundle $\mathbb{C}\otimes\Lambda(\mathbb{R}^4)$.  This bundle contains sixteen complex components, and the Dirac-K\"ahler equation on it is equivalent to the classical Dirac equation for four Dirac bispinors.  On the other hand, for this equation exists a doubler-free so-called ``staggered'' or Kogut-Suskind discretization (\cite{Kogut}, see also \cite{Striker}).  Instead of sixteen doublers of the naive discretization this gives a lattice equation for four Dirac bispinors.   Another solution is to suppress the doublers using some additional terms which give all the doublers except one a large mass.  The most famous one is the Wilson mass term \cite{Wilson}, but many different versions of such mass terms have been proposed.

In this paper we consider another variant of geometric fermions.  Instead of the bundle $\mathbb{C}\otimes\Lambda(\mathbb{R}^4)$ we use the three-dimensional exterior bundle $\Lambda(\mathbb{R}^3)$ and a direction-dependent complex structure on it.  The four-dimensional Dirac equation appears as an evolution equation on this bundle.  The time $t$ is only a parameter of dynamics without geometric meaning.
This representation of the Dirac equation describes already a single bispinor.

The main advantage of the geometric Dirac-K\"ahler fermions remains --- there exists a doubler-free ``staggered'' discretization of the continuous Dirac equation.  But now it
gives a doubler-free ``staggered'' discretization of a single bispinor.  We find also a remarkable, more symmetric discretization for an octet of bispinors related with the bundle $(\Lambda\otimes\Lambda)(\mathbb{R}^3)$.

Using the standard Wilson approach to lattice gauge fields becomes problematic because the complex structure does not act pointwise.  But to put the octet representation of $SU(3)$ used in QCD on the lattice we do not need the complex structure: It is equivalent to a real representation.  As a consequence we obtain a doubler-free discretization for lattice QCD.

The discretization presented here is closely related with the ``space-geometric'' interpretation of standard model fermions as sections of the bundle $(T\otimes\Lambda\otimes\Lambda)(\mathbb{R}^3)$, or, on curved background with ADM decomposition $M^4\cong M^3\oplus\mathbb{R}$, the bundle $(T\otimes\Lambda\otimes\Lambda)(M^3)$ made in \cite{SMbundle}.

\section{Space-geometric representation of spinors}

The classical Dirac algebra is defined by the operators $\gamma^i$ with commutation relations $\{\gamma^i,\gamma^j\}=\eta^{ij}$ --- the Clifford algebra $Cl(1,3)$.  But there are also some other operators which play an important role in the standard model: The complex structure, which is defined by the operator $i$, and a real structure defined by the operator $C$, which is necessary to define Majorana mass terms.  The commutation relations of this ``extended Dirac algebra'' $\{\gamma^\mu, i, C\}$ are: $\{\gamma^\mu,\gamma^\nu\}=2\eta^{\mu\nu}$, $[i,\gamma^\mu]=0$, $[C,\gamma^\mu]=0$, $\{C,i\}=0$, $C^2=1$, $i^2=-1$.

The three-dimensional or space-geometric representation of spinors based on the following isomorphism:

\begin{theorem} \label{t1}
The ``extended Dirac algebra'' $\{\gamma^\mu, i, C\}$ is isomorph to the Clifford algebra $Cl(3,3)$.  The Clifford algebra generators are the $\gamma^i$ for $i>0$ and the three operators $\beta^i$ defined by

\begin{eqnarray*}
\beta^1&=&C,\\
\beta^2&=&Ci,\\
\beta^3&=&\gamma^5=-i\gamma^0\gamma^1\gamma^2\gamma^3.
\end{eqnarray*}
\end{theorem}

Proof: An explicit check proves that the operators $\gamma^i,\beta^i$ fulfil the properties required for the generators of $Cl(3,3)$ (they anticommute, $(\beta^i)^2=-(\gamma^i)^2=1$).  We have already expressed the generators of $Cl(3,3)$ in terms of the extended Dirac algebra $\{\gamma^\mu, i, C\}$.  The reverse expressions can be easily found:

\begin{eqnarray}
C &=& \beta^1\\
i &=& \beta^1\beta^2\\
\gamma^0 &=& \beta^1\gamma^1\beta^2\gamma^2\beta^3\gamma^3
\end{eqnarray}

It remains to prove that there are no other relations in  $\{\gamma^\mu, i, C\}$.  This can be done by an explicit computation of the dimension of the extended Dirac algebra.  We have six generators  $\{\gamma^\mu, i, C\}$, for each pair of generators we have one commutation resp. anticommutation rule, the square of all generators is $\pm 1$, and there are no other independent relations.  Therefore an arbitrary monom may be transformed into a monom where every generator appears at most once.  There are $2^6$ such monoms, which gives the dimension of $Cl(3,3)$. qed.

The Clifford algebra $Cl(3,3)$ acts in a natural way on the three-dimensional Clifford bundle $Cl(\mathbb{R}^3)$.  This bundle is equivalent to the exterior bundle $\Lambda(\mathbb{R}^3)$ which consists of inhomogeneous differential forms $\varphi=\varphi_\kappa(x) (dx^1)^{\kappa_1}(dx^2)^{\kappa_2}(dx^3)^{\kappa_3}$, $\kappa=(\kappa_1,\kappa_2,\kappa_3)\in(\mathbb{Z}_2)^3$.  The operators $\gamma^i$ act as in standard Hodge theory and define the Hodge theory Dirac operator

\begin{equation}
\label{D3} D_3 = i \gamma^i \partial_i
\end{equation}

(see, for example, \cite{Pete}).  The action of the operators $\beta^i$ follows from the following formula:

\begin{equation}
\label{betadef} (\beta^i\gamma^i\varphi)_\kappa(x) = (-1)^{\kappa_i} \varphi_\kappa(x)
\end{equation}

It is not hard to see the following:

\begin{theorem}
The representation of the extended Dirac algebra $\{\gamma^\mu, i, C\}$ on $\Lambda(\mathbb{R}^3)$ defined by (\ref{D3}), (\ref{betadef}) and theorem \ref{t1} is equivalent to the standard representation.
\end{theorem}

Theorem \ref{t1} gives now the following four-dimensional Dirac operator:

\begin{equation}
\label{D4} D_4 = i\gamma^0 \partial_0 + D_3
\end{equation}

The definition of the Lagrange formalism simplifies if we use the following variant of the Dirac operator:

\begin{equation}
\label{D} D = -i\gamma^1\gamma^2\gamma^3 D_4 = \iota \partial_t + \iota \alpha^i \partial_i
\end{equation}

where  $\iota=\beta^1\beta^2\beta^3$, $\iota^2=-1$, $\{\iota,\gamma^i\}=0$, $[\iota,\alpha^i]=0$.  The equivalence is obvious: $D\varphi=0$ iff $D_4\varphi=0$.  Now we can use the standard Euclidean scalar product on $\Lambda(\mathbb{R}^3)$ to define the Langrange formalism as

\begin{equation}
\label{Lcontinuous}L = (\varphi D \varphi)
\end{equation}

The Dirac operator considered here should not be mingled with the four-dimensional Dirac-K\"ahler operator defined on the spacetime bundle $\mathbb{C}\otimes\Lambda(\mathbb{R}^4)$ which has real dimension 32.  It defines an evolution equation on the spatial bundle $\Lambda(\mathbb{R}^3)$ with real dimension 8.  Time is only a parameter of dynamics and has no geometric interpretation. Manifest Lorentz invariance is lost.  Only the equivalence theorem proves that Lorentz invariance will be recovered for observable effects.

This representation also violates manifest symmetry in space:  The complex structure $i=\beta^1\beta^2$ as well as the operator $\gamma^5=\beta^3$ depend on a preferred direction in space.  The introduction of three generations of fermions associated with the three spatial directions may be used to get rid of this asymmetry.

A similar "space Clifford" representation of the Dirac equation has been considered by Daviau \cite{Daviau}, see also \cite{Fauser}.

A comparison of space-geometric spinors with standard spinors can be found in \cite{SMbundle}. It is shown there that on Minkowski background we have no observable differences, but on curved background we obtain differences which are observable in a graviational analogon of the Bohm-Aharonov experiment.  

\section{Lattice theory for the three-dimensional Dirac operator}

Let's consider at first the naive central differences discretization of the Dirac operator $D$ on $\Lambda(\mathbb{R}^3)$.  We use a regular rectangular lattice with lattice functions $\varphi\in\Omega:\mathbb{Z}_2^3\otimes\mathbb{Z}^4\to\mathbb{R}$ denoted by $\varphi_\kappa(n)\in\mathbb{R}$, $\kappa=(\kappa_1,\kappa_2,\kappa_3)\in(\mathbb{Z}_2)^3$, $n=(n_0,n_1,n_2,n_3)\in\mathbb{Z}^4$. Using the denotations $h_\mu=(\delta_{0\mu},\delta_{1\mu},\delta_{2\mu},\delta_{3\mu})$ we define

\begin{eqnarray}
(t^\mu_\pm\varphi)_\kappa(n) &=& \varphi_\kappa(n\pm h_\mu)\\
\Delta_\mu &=& \frac{1}{2}(t^\mu_+ - t^\mu_-)\\
\label{Dnaive} D_h &=& \iota \Delta_0 + \iota\alpha^i\Delta_i
\end{eqnarray}

The Lagrange formulation of the lattice theory can be given in the same form as (\ref{Lcontinuous}), but with the lattice version (\ref{Dnaive}) of the Dirac operator $D$:

\begin{equation}
\label{Lnaive}L = (\varphi D_h \varphi)
\end{equation}

\subsection{Decomposition into sixteen staggered fermions}

A well-known problem of lattice gauge theory --- the so-called ``fermion doubling problem'' --- is that the naive central differences discretization of the classical Dirac equation does not describe a single Dirac fermion, but sixteen.   This doubling problem appears in our naive discretization (\ref{Dnaive}) too.  As usual, we obtain sixteen Dirac fermions instead of one. But in our discretization these sixteen doublers decompose in a quite simple way.

\begin{theorem} There exists a decomposition

\[
\Omega\cong\sum_\lambda \Omega^\lambda,
\]

$\lambda=(\lambda_0,\lambda_1,\lambda_2,\lambda_3)\in \mathbb{Z}^4$, $\lambda_i=\pm 1$ so that for $\varphi=\sum \varphi_\lambda$, $\varphi_\lambda\in\Omega^\lambda$

\[
D_h \varphi = 0 \hspace{1cm} \Rightarrow \hspace{1cm} D_h \varphi_\lambda = 0
\]

and the subspaces $\Omega^\lambda\subset\Omega$ are realized as staggered sublattices.
\end{theorem}

To see this let's consider the following lattice operators:

\begin{eqnarray}
(\epsilon_0 \varphi)_\kappa(n) &=& (-1)^{n_0+n_1+n_2+n_3} \varphi_\kappa(n)\\
(\epsilon_i \varphi)_\kappa(n) &=& (-1)^{k_i+n_i} \varphi_\kappa(n)
\end{eqnarray}

These four operators $\epsilon_\mu$ commute with each other and fulfil $\epsilon_\mu^2=1$.  This allows to define common eigenspaces with eigenvalues $\pm 1$.  Thus, we define the $\Omega^\lambda$ as eigenspaces $\epsilon_\mu\varphi=\lambda_\mu\varphi$.  It is easy to verify that the $\Omega^\lambda$ define staggered sublattices:  They contain only the component defined by $(-1)^{k_i+n_i}=\lambda_i$ on nodes which fulfil the property $(-1)^{n_0+n_1+n_2+n_3}=\lambda_0$.

Now, the $\epsilon_\mu$ also anticommute with $D_h$.  This gives the property
$D_h: \Omega^\lambda \to \Omega^{-\lambda}$.  As a consequence,
from $D_h\varphi=0$ follows $D_h\varphi_\lambda=0$ for all $\lambda$.

The decomposition on the level of lattice components suggests to use one of the eigenspaces, say $\Omega^{(1,1,1,1)}$, as a staggered discretization of a single Dirac fermion.

A comparison of such a single staggered fermion with the standard staggered Dirac-K\"ahler fermions seems interesting:  In the classical staggered fermion concept \cite{Kogut} there remain four doublers (more accurate, already the continuous Dirac-K\"ahler equation describes four Dirac bispinors, it's staggered discretization is doubler-free as a discretization of the Dirac-K\"ahler equation but nonetheless describes four bispinors).  Instead, in our approach we have no doublers, but a single Dirac fermion (eight real components).  On the other hand, we have to pay for this with two disadvantages:  We have no longer manifest, explicit relativistic symmetry, but a special role of the time coordinate.  On curved background, such a special role for a time coordinate requires an additional structure --- an ADM decomposition.

The other problem is the complex structure.  First, it depends on a direction in space. This dependence disappears if we introduce three generations of fermions and associate the three generations with directions in space (see \cite{SMbundle}).  A more serious problem is that the complex structure on the lattice is not pointwise: real and imaginary parts of the four complex components of a bispinor are located on different lattice nodes.  Especially there is no pointwise operator for multiplication with $i$ on our staggered fermion.  Now, multiplication with $i$ is obviously an important operation in the standard model.  Even if we get rid of it in the pure Dirac equation using the form (\ref{D4}) or (\ref{D}), they reappear in the interaction terms with gauge fields as well as in the mass terms.  Fortunately for mass terms it seems sufficient to fix some non-pointwise replacement for $i$.  For gauge fields, the non-pointwise character of multiplication with $i$ seems more problematic.

\subsection{The geometric octet}

To define the standard model (inclusive right-handed neutrinos), we have to consider three octets of Dirac fermions.  Now, a straightforward way to do define an octet would be to use eight copies of the staggered fermion defined by $\Omega^{(1,1,1,1)}$. But there is a much simpler and more natural way to obtain an octet of fermions which we name {\em geometric octet\/}:  We use the eigenspace $\hat{\Omega}\subset\Omega$ defined by $\epsilon_0\varphi=\varphi$ which consists of all components on "even" nodes. The lattice space $\hat{\Omega}$ has not only a much simpler definition (we don't even have to define the operators $\epsilon_i, i>0$ to define $\hat{\Omega}$), but also a larger symmetry group: While on different lattice nodes in $\Omega^{(1,1,1,1)}$ we have different components $\varphi_\kappa$, all lattice nodes in $\hat{\Omega}$ are of the same type: They contain all eight components $\varphi_\kappa$.

Thus, from point of view of symmetry between lattice nodes, the geometric octet of Dirac fermions seems much more natural than a single Dirac fermion.  Let's consider its geometric nature:  The eight doublers $\Omega^{(1,i,j,k)}$ in this octet are connected with each other by translations in space.  The components of the continuous limit of the geometric octet therefore should be connected by infinitesimal translations. This gives transformational properties as for the geometric bundle $(\Lambda\otimes\Lambda)(\mathbb{R}^3)$ instead of $\Lambda(\mathbb{R}^3)^8$.\footnote{Of course, as long as we consider only the Minkowski background, as we do in this paper, there is no physical difference between $(\Lambda\otimes\Lambda)(\mathbb{R}^3)$ and $\Lambda(\mathbb{R}^3)^8$. The difference becomes important on curved background.  The consideration of discretizations for curved background we leave to future research.}

There is one problem with the reduction $\Omega\to\hat{\Omega}$ --- the naive Lagrange formulation (\ref{Lnaive}) is not well-defined on $\hat{\Omega}$.  But there is a simple solution --- instead of (\ref{Lnaive}) we can use the following Lagrangian:

\begin{equation}
\label{Lshift}L = (\varphi t^3 D_h \varphi)
\end{equation}

where the translation $t^3$ defines a preferred direction in space. This requires the same type of symmetry breaking as the choice of the complex structure $i=\beta^1\beta^2$, and we can get rid of it if we introduce three fermion generations (see \cite{SMbundle}).

Below we focus our interest on the ``geometric'' octet.

\section{Lattice gauge theory}

Our lattice Dirac equation for the geometric octet may be connected with gauge fields following the classical Wilson scheme for lattice gauge fields, even with some simplifications. In the Wilson approach, the lattice gauge field for a gauge group $G$ is described by gauge transformations $U_\mu(n)\in G$ defined on edges (links) of the lattice.  The lattice gauge transformations are defined by functions $g(n)\in G$ on the nodes with values in the gauge group $G$.  In comparison with standard lattice fermions we have no pointwise operation of multiplication with $i$.  But if the representation $T: G\to gl(8)$ which acts on the space spanned by the $\varphi_\kappa(n)$ is real, we can nonetheless use the standard way to define a gauge-invariant interaction Lagrangian --- to replace the difference operators in the Lagrangian by covariant difference operators $\Delta_\mu \to \tilde{\Delta}_\mu$:

\begin{equation} (\tilde{\Delta}_\mu\varphi)_\kappa(n)=\frac{1}{2}(T(U_\mu(n))\varphi_\kappa(n+h_\mu)-T(U_\mu^{-1}(n-h_\mu,\mu))\varphi_\kappa(n-h_\mu))
\end{equation}

The restriction to pointwise gauge transformations (and the resulting restriction to real representations) is is not only prescribed by the standard Wilson approach. It also allows to preserve the reduction $\Omega\to\hat{\Omega}$ and, therefore, allows to define the gauge action not only on the original sixteen doublers $\Omega$ but also on the geometric octet $\hat{\Omega}$. This gives the Lagrangian

\begin{equation}
L = (\varphi t^3 \tilde{D}_h \varphi) + L_{Wilson}(U_\mu(n))
\end{equation}

for the gauged Dirac operator $\tilde{D_h}= \iota \tilde{\Delta}_0 + \iota\alpha^i\tilde{\Delta}_i$.

\subsection{QCD on the geometric octet}

Let's consider now the specific properties of local gauge actions on the geometric octet.  To define QCD on the geometric octet we need the appropriate octet representation of the Lie algebra $su(3)$ in terms of pointwise operators which commute with the Dirac operator $D_h$.  For this purpose, let's consider the following operator algebra:

\begin{eqnarray}
\beta_i &=& \beta^i\\
\gamma_i&=&\beta_i\epsilon_i
\end{eqnarray}

We have $\beta_i^2=1$, $\gamma_i^2=-1$, and all $\beta_i,\gamma_i$ anticommute, thus, it is the Clifford algebra $Cl(3,3)\cong gl(8)$. The compact subalgebra $o(8)\subset gl(8)$ is generated by products $g$ of the generators which fulfil $g^2=-1$.

Moreover, we have simple commutation relations with the Dirac operator:  $[\beta_i,D_h]=0$, $\{\gamma_i,D_h\}=0$.  If we require $[g,D_h]=0$, we have to restrict ourself to monoms with even number of factors $\gamma_i$.  This restriction is equivalent to $[g,\hat{i}]=0$ with $\hat{i}=\beta_1\beta_2\beta_3$, $\hat{i}^2=-1$.  The subalgebra of $o(8)$ which fulfills this restriction is equivalent to $u(4)$, where $\hat{i}$ plays the role of the related complex unit.

It should be noted that this is a real eight-dimensional representation of $u(4)$.  A complex unit $i$ which acts on single fermions $\Omega^{(1,i,j,k)}$ has not been used here.  Such a complex unit $i$ should be distinguished from the real operator $\hat{i} \in gl(8)$ which does not preserve the decomposition into fermions $\Omega^{(1,i,j,k)}$.

As a real representation, this octet representation of $u(4)$ would be irreducible.  But as a complex octet representation it is reducible: We have an involution operator $i\hat{i}$, $(i\hat{i})^2=1$,  $[g,i\hat{i}]=0$ for $g\in u(4)$.  This allows to define two invariant subspaces with projectors $P^\pm=\frac{1}{2}(1\pm i\hat{i})$. On these four-dimensional subspaces $\hat{\Omega}_\pm=P^\pm\hat{\Omega}$ we have $i=\mp\hat{i}$.\footnote{Note that on the lattice such a decomposition would require a lattice analogon of $i$ with exact properties $i^2=-1$, $[i,\hat{i}]=0$, $[i,\tilde{D}_h]=0$ which does not exist.  Thus, this decomposition will not be exact on the lattice. It appears only in the continuous limit.}  The observable quarks and leptons will be, therefore, linear combinations of pairs of staggered fermions
$\Omega^{(1,i,j,k)}$ and $\Omega^{(1,-i,-j,-k)}$ defined by the projectors $P^\pm$.  It seems natural to identify the pairs $\Omega^{(1,i,j,k)}\oplus\Omega^{(1,-i,-j,-k)}$ with electroweak doublets.

Thus, we obtain two times the defining four-dimensional representation of $u(4)$.  Restricted to $su(3)\subset u(4)$ this gives two times the defining representation of $su(3)$ and two times a singlet representation of $su(3)$, as required for the QCD action.  At least as long as we do not consider weak interaction the two subspaces $\hat{\Omega}_\pm$ are independent. Therefore the different choice of the sign of the complex structure is a question of convention.

It remains to define an appropriate restriction from $u(4)$ to $su(3)$.  For this purpose we have the interesting symmetric operator

\begin{equation}
\sigma=\beta_1\gamma_1\beta_2\gamma_2+\beta_2\gamma_2\beta_3\gamma_3+\beta_3\gamma_3\beta_1\gamma_1
\end{equation}

which defines baryon resp. lepton charge in a natural way.  The restriction $[g,\sigma]=0$ defines the subalgebra $u(1)\oplus u(3)\subset u(4)$.  Its generators are:

\begin{equation}
\hat{i} = \beta_1\beta_2\beta_3, \hspace{0.5cm}
\delta_i = \beta_i\gamma_j\gamma_k, \hspace{0.5cm}
\omega_i = \beta_j\beta_k-\gamma_j\gamma_k, \hspace{0.5cm}
\eta_i = \gamma_i(\beta_j\gamma_j-\beta_k\gamma_k),
\end{equation}

with $1\le i\le 3$, $j=i+1 \mbox{ mod } 3$, $k=i+2 \mbox{ mod } 3$.
The two-dimensional center of this algebra is generated by the generators $\hat{i}$, $\delta = \delta_1+\delta_2+\delta_3$.  We obtain $su(3)$ as the remaining factor algebra, with the eight generators $\omega_i$, $\eta_i$, $\delta_1-\delta_2$, $\delta_2-\delta_3$.

\subsection{The mass terms}\label{mass}

To put various mass terms on the lattice we need lattice versions of the complex unit $i=\beta^1\beta^2$ as well as of the operators $C=\beta^1$, $\gamma^5=\beta^3$.  These have to be operators which leave the subspaces $\Omega^\lambda$ invariant.  Because they are transformations of different components $\varphi_\kappa$, which are located on different nodes in the staggered subspaces $\Omega^\lambda$, such operators cannot be pointwise operators.   As a consequence, it is not possible to preserve all important exact properties of the continuous operators $i, C, \gamma^5$ on the lattice.  The question is which of the exact properties are the more important.  For example, a variant for $i$ which commutes with $D_h$ is:

\begin{equation}\label{i1}
(i_h\varphi)_\kappa(n)=(\beta^1\beta^2\varphi)_\kappa(n+h_1+h_2).
\end{equation}

But this operator does not define a complex structure, $(i_h)^2\neq -1$ but includes a shift.  We can, instead, preserve the property $i^2=-1$, for example with:

\begin{equation}\label{i2}
(i_h\varphi)_\kappa(n)=(\beta^1\beta^2\varphi)_\kappa(n+(-1)^{n_1}h_1+(-1)^{n_2}h_2)
\end{equation}

In this case $[i_h,D_h]\neq 0$.  Similar variants exist for the operators $C$, $\gamma^5$.  Which choices for these operators are optimal for the implementation of mass terms has to be left to future research.

\section{Discussion}

We have defined a doubler-free gauge-invariant lattice gauge theory for the $SU(3)$ octet action of QCD.  Such a discretization is in itself quite useful for applications in lattice QCD:  In comparison with the standard approach to lattice QCD we do not have to waste time and memory for the doublers, and we do not have to compute the additional terms to suppress them.

But this discretization seems interesting not only because of these applications.  It has been found by consideration of a ``space-geometric'' interpretation of the whole fermionic sector of the standard model as the bundle $(T\otimes\Lambda\otimes\Lambda)(\mathbb{R}^3)$ proposed in \cite{SMbundle}. The vector index describes the three fermion generations, and the bundle $(\Lambda\otimes\Lambda)(\mathbb{R}^3)$ is the large distance limit of our geometric octet.  The direction-dependent structures of our octet (the complex structure $i=\beta^2\beta^3$, use of $t^3$ in (\ref{Lshift})) may be combined into direction-independent structures if we consider all three generations and associate these generations with different preferred directions in space.

On the other hand, there is no simple way to recover manifest relativistic symmetry.  In the space-geometric interpretation time is only a non-geometric parameter of evolution. Geometric interpretation is only three-dimensional. This is in conflict with the four-dimensional spacetime interpretation of the gravitational field, but in natural correspondence  with the ADM decomposition --- a foliation of spacetime into space and time $M^4\cong M^3\oplus \mathbb{R}$.  In an ADM decomposition the gravitational field is described by a positive-definite three-metric, a three-dimensional vector field, and a positive scalar field.  In \cite{SMbundle} we propose to interpret standard model fermions as sections of the bundle $(T\otimes\Lambda\otimes\Lambda)(M^3)$.

If this discretization really has some fundamental importance, then there should be also a natural possibility to describe electroweak interaction on the geometric octet.  The key for understanding chiral gauge fields may be a better understanding of the necessarily non-pointwise lattice approximations of $i$, $C$, and $\gamma^5$.  Note, for example, that the use of an $i_h$ which preserves $(i_h)^2=-1$ like (\ref{i2})
defines a decomposition into two lattice subspaces $\hat{\Omega}^\pm_h$ which is not exactly preserved by $D_h$.  A decomposition into exact mass eigenstates on the lattice will give different subspaces which will not be left invariant by the QCD gauge fields. They will coincide only in the continuous limit.  What will be the highest order correction terms?  It will be some interaction between the mass eigenstates inside the electroweak doublets $\Omega^{(1,i,j,k)}\oplus\Omega^{(1,-i,-j,-k)}$.

Maybe for some appropriate choice of the mass terms this discretization effect gives  electroweak interaction or something sufficiently close to it?  This hypothesis we leave to future research.

\end{document}